\documentclass[aps,prd,nofootinbib,showkeys,preprint,floatfix]{revtex4}

\usepackage{amssymb}
\usepackage{amsmath}
\usepackage{amsfonts}
\usepackage{graphicx}
\usepackage{multirow,rotating}
\usepackage{color}
\usepackage{hyperref}

\newcommand {\eq} [1] {eq.~(\ref{eq:#1})}

\newcommand{\AddrAHEP}{
  {\it AHEP Group, Instituto de F\'{\i}sica Corpuscular --
    C.S.I.C./Universitat de Val{\`e}ncia \\
    Edificio de Institutos de Paterna, Apartado 22085,
  E--46071 Val{\`e}ncia, Spain}}

\newcommand{\AddrWur}{
Institut f\"ur Theoretische Physik und Astronomie, 
Universit\"at W\"urzburg\\
Am Hubland, 
97074 W\"urzburg, {Germany}}

\newcommand{\AddrBonn}{
Physikalisches Institut der Universit\"at Bonn \\
Nussallee 12, 53115 Bonn, Germany
}

\def\gsim{\raise0.3ex\hbox{$\;>$\kern-0.75em\raise-1.1ex\hbox{$\sim\;$}}}
\def\lsim{\raise0.3ex\hbox{$\;<$\kern-0.75em\raise-1.1ex\hbox{$\sim\;$}}}

\newcommand{\Ds}{\Delta m^2_{\odot}}
\newcommand{\Da}{\Delta m^2_{\textsc{A}}}

\begin{document}

\preprint{IFIC/11-59,  Bonn-TH-2011-15}  

\title{Hefty MSSM-like light Higgs in extended gauge models}

\author{M. Hirsch} \email{mahirsch@ific.uv.es}\affiliation{\AddrAHEP}
\author{M. Malinsk\'y}\email{malinsky@ific.uv.es}\affiliation{\AddrAHEP}
\author{W. Porod} \email{porod@physik.uni-wuerzburg.de}\affiliation{\AddrWur \\ \AddrAHEP}
\author{L.~Reichert} \email{reichert@ific.uv.es}\affiliation{\AddrAHEP}
\author{F. Staub}\email{fnstaub@th.physik.uni-bonn.de}\affiliation{\AddrBonn {\rm and} \\ \AddrWur}



\begin{abstract}
It is well known that in the MSSM the lightest neutral Higgs $h^{0}$ must be, 
at the tree level, lighter than the $Z$ boson and that the loop corrections 
shift this stringent upper bound up to about 130~GeV.
Extending the MSSM gauge group in a suitable way, the new Higgs sector dynamics 
can push the tree-level mass of  $h^{0}$ well above the tree-level MSSM limit 
if it couples to the new gauge sector.
This effect is further pronounced at the loop level and $h^{0}$ masses in the 
140 GeV ballpark can be reached easily. We exemplify this for a sample setting 
with a low-scale $U(1)_{R}\times U(1)_{B-L}$ gauge symmetry in which 
neutrino masses can be implemented via the inverse seesaw mechanism.
\end{abstract}

\maketitle

\section{Introduction}
\label{sec:intro}

With the recent start of the LHC the hunt for the Higgs has
recommenced and new limits for the Standard Model (SM) Higgs bosons
have been obtained excluding various mass ranges between 145~GeV and
470~GeV \cite{ATLAStalk,CMStalk}. At the same time, a slight excess of
events in the mass window at around 140 GeV has been observed.  Unlike
in the SM, there is a stringent tree-level upper bound on the mass of
the lightest CP-even Higgs state in the Minimal Supersymmetric
extension of the Standard Model (MSSM) $m_{h^0} \le {\rm
  Min}(m_A,m_{Z})$ where $m_{A}$ denotes the mass of the
pseudoscalar Higgs and $m_{Z}$ is the mass of the SM $Z$-boson.
This comes from the fact that supersymmetry links the MSSM Higgs
self-interactions to the gauge couplings (via the so-called
$D$-terms), thus making the scalar potential of the theory rather
rigid. It is well known 
\cite{Ellis:1990nz,Ellis:1991zd,Okada:1990vk,Haber:1990aw} 
that radiative corrections are important in the Higgs sector of
the MSSM.
However, even at 2-loop and 3-loop level one obtains on upper
bound of $m_{h^0} \lsim 130$ GeV
\cite{Hempfling:1993qq,Heinemeyer:1998jw,Heinemeyer:1998np,Espinosa:1999zm,Degrassi:2001yf,Brignole:2002bz,Brignole:2001jy,Carena:2000dp,Degrassi:2002fi,Allanach:2004rh,Harlander:2008ju,Kant:2010tf} for SUSY particles
below roughly $m_{SUSY} \sim 2$ TeV and a top quark mass of about 173
GeV.  If the above hint for a Higgs in the range of 140 GeV turns out
to be correct, this would essentially rule out the MSSM, except
perhaps for split-SUSY-like scenarios \cite{Giudice:2004tc} where the
squarks and sleptons are typically pushed far above the TeV scale.
One popular way to resolve this issue is to add additional Higgs
fields, e.g., the NMSSM-like singlet(s) \cite{Ellwanger:2009dp}, so
that extra $F$-term contributions to the scalar potential lift the
Higgs mass. Another option is to enhance the $D$-terms by employing
extended gauge symmetries \cite{Haber:1986gz,Drees:1987tp,Cvetic:1997ky,Zhang:2008jm,Ma:2011ea}.

Besides the Higgs mass puzzle (assuming the Higgs boson exists at all) there is yet another eminent mass-related 
riddle in the  particle physics, namely, why neutrinos are so much
lighter than all other matter particles. In the ``standard'' seesaw picture  \cite{Minkowski:1977sc,MohSen,Schechter:1980gr,Cheng:1980qt} this is  attributed to a new very-high-energy dynamics (often in the vicinity of the GUT scale) which, at low energies, exhibits itself as a dimension-five effective operator 
giving the SM neutrinos a lepton-number violating Majorana mass \cite{Weinberg:1979sa,Weinberg:1980bf}.
On the other hand, there is nothing really fundamental about such high-energy realizations of the seesaw mechanism as, in principle, variants of seesaw can be  implemented at virtually any mass scale; inverse seesaw proposed in 
\cite{Mohapatra:1986bd} or the linear seesaw of  
\cite{Akhmedov:1995ip} are just two examples. Such schemes are naturally realized in the class of  left-right symmetric extensions of the SM \cite{Malinsky:2005bi,Dev:2009aw} based on the popular $SO(10)$ breaking chains
\begin{eqnarray}
SO(10) &\to &  SU(4)_{C} \times SU(2)_R\times SU(2)_L\to 
SU(3)_c\times SU(2)_L \times U(1)_Y \\
SO(10) & \to & SU(3)_c\times SU(2)_R\times SU(2)_L \times U(1)_{B-L}
 \to SU(3)_c\times SU(2)_L \times U(1)_Y 
\end{eqnarray}
which, however, often call for further extension of the matter sector in order to maintain the  near-perfect gauge coupling unification of the 
MSSM \cite{Majee:2007uv}. A potential problem in this context
is that the gauge couplings can easily become non-perturbative well below the GUT scale as matter particles always yield a positive contribution
to the corresponding beta functions~\cite{Kopp:2009xt}. 
Nevertheless, one can still devise models where an extended gauge symmetry can remain unbroken down to almost the  
electroweak scale and, at the same time,  a perturbative unification of the gauge couplings 
 at around $10^{16}$ GeV is retained \cite{DeRomeri:2011ie}. 

In such extended models the MSSM Higgs bosons are often 
 charged also under the additional group factor(s). Hence, extra $D$-terms contributing
to the masses of the neutral Higgs bosons are naturally supplied. This implies that
the  Higgs boson, which is mainly the MSSM $h^0$,
can have a mass above the $Z$-boson mass already at the tree level.
Moreover, the additional Higgs fields needed to break the
extended gauge symmetry mix with the usual two Higgs doublets
of the MSSM, thus affecting the phenomenology of the Higgs sector. In particular, \hyphenation{par-ti-cu-lar}
the couplings of the mainly $SU(2)_L$-doublet-like Higgs bosons
to the SM particles can get reduced and one
can find parameter regions where the lightest MSSM-like Higgs boson $h^0$
decays into two of the additional Higgs bosons, as they can be
very light  without violating
existing experimental bounds \cite{Barate:2003sz}. 
 
In this letter we  exemplify this basic mechanism at the Higgs sector of a simple extension of the SM
featuring a $U(1)_R \times U(1)_{B-L}$ gauge symmetry 
which gets broken to  $U(1)_Y$ at energies close to the electroweak scale.
The $U(1)_R$ factor can be viewed as,  e.g., a remnant of a complete gauged $SU(2)_R$ symmetry that can be restored at higher energies, thus facilitating a possible embedding into a full GUT based,  for instance, on an $SO(10)$ gauge symmetry~\cite{Malinsky:2005bi}.
To this end, we complement the existing literature in several aspects; namely, by performing a complete one-loop analysis of the light Higgs sector and by checking that the light Higgs phenomenology is fully consistent with the current data by inspecting carefully all the relevant  constraints from colliders (in particular, those coming from the LEP and LHC searches) and lepton flavour violation ($\mu \to e \gamma$). In particular, we consider not only shifts in the masses of the lightest Higgs CP-even eigenstates but also the changes in their character, i.e., the amount of the $SU(2)_{L}$ doublet components within, and their implications.

In the next section we present the details
of the model focusing namely on its extended Higgs sector. Working out the relevant mass matrices we argue that the CP-even mass eigenstate most similar to the  
lightest MSSM Higgs boson $h^0$ can have a mass above 100~GeV
already at the tree level.
We also briefly discuss the one-loop corrections to the tree-level situation. 
In Section \ref{sec:numerics} we present results of a dedicated numerical analysis where the complete one-loop 
corrections to the Higgs sector were taken into account. Finally,  we draw our conclusions in Section \ref{sec:conc}.

\section{The model and its Higgs sector}
\label{sec:model}

\begin{table}
\begin{center} 
\begin{tabular}{|c|c|c|c|c|} 
\hline \hline 
\mbox{}\;\;\;\;\;\mbox{}& \; Superfield\;  & \; $SU(3)_c\times SU(2)_L\times U(1)_R\times U(1)_{B-L}$\;& \; Generations \; \\ 
\hline 
\multirow{7}{*}{\begin{sideways}Matter\end{sideways}}& \(\hat{Q}\)  & \(({\bf 3},{\bf 2},0,+\frac{1}{6}) \) & 3 \\ 
&\(\hat{d^c}\) & 
\(({\bf \overline{3}},{\bf 1},+\frac{1}{2},-\frac{1}{6}) \)& 3 \\ 
&\(\hat{u^c}\) & 
\(({\bf \overline{3}},{\bf 1},-\frac{1}{2},-\frac{1}{6}) \)& 3 \\ 
&\(\hat{L}\)  & \(({\bf 1},{\bf 2},0,-\frac{1}{2}) \) & 3 \\ 
&\(\hat{e^c}\) & \(({\bf 1},{\bf 1},+\frac{1}{2},+\frac{1}{2}) \) & 3 \\ 
&\(\hat{\nu^c}\) & \(({\bf 1},{\bf 1},-\frac{1}{2},+\frac{1}{2}) \) & 3 \\ 
&\(\hat S\)& \(({\bf 1},{\bf 1},0,0) \) & 3 \\ 
\hline
\multirow{4}{*}{\begin{sideways}Higgs\end{sideways}}&\(\hat{H}_u\)  & \(({\bf 1},{\bf 2},+\frac{1}{2},0) \) & 1 \\ 
&\(\hat{H}_d\)  & \(({\bf 1},{\bf 2},-\frac{1}{2},0) \) & 1 \\ 
&\(\hat{\chi}_R\)  & \(({\bf 1},{\bf 1},+\frac{1}{2},-\frac{1}{2}) \) & 1 \\ 
&\(\hat{\bar{\chi}}_R\)  & \(({\bf 1},{\bf 1},-\frac{1}{2},+\frac{1}{2})\) & 1 \\ 
\hline \hline
\end{tabular} 
\end{center} 
\caption{\label{tab:fc}The Matter and Higgs sector field content of the $U(1)_{R}$ model under consideration. Matter generation indices have been suppressed. The $\hat S$ superfields are included to generate neutrino masses 
via the inverse seesaw mechanism. Under matter parity, the matter fields are odd while the Higgses are even. The additive charges were conveniently normalized in such a way that  $Y=T_{R}+B-L$ and $Q=T_{L}^{3}+Y$.}
\end{table}

We shall consider a sample model based on the $SU(3)_{c}\times SU(2)_L\times U(1)_R \times U(1)_{B-L}$ gauge group 
 which can emerge, e.g.,  in
a class of  $SO(10)$ GUTs broken along the ``minimal'' left-right symmetric chain 
\begin{equation}
SO(10) \to  SU(3)_c\times SU(2)_L\times SU(2)_R \times U(1)_{B-L}
\to
   SU(3)_c\times SU(2)_L\times U(1)_R \times U(1)_{B-L},\nonumber
\end{equation}
as advocated, for instance, in \cite{Malinsky:2005bi}. The main virtue of this setting is that, even in the minimally fine-tuned version, an MSSM-like gauge unification is perfectly compatible with a $U(1)_R \times U(1)_{B-L}$ stage stretching down to a few TeV. 
Renormalization
group evaluation usually lead to $U(1)$ mixing effects
 \cite{delAguila:1988jz}
if more than
one $U(1)$ factor is present which not only introduces additional
couplings in the gauge sector but also affects the evolution of
the soft SUSY breaking parameters \cite{Fonseca:2011vn}. 
However, it turns out that for the model under study the corresponding
effects are small \cite{followup}. Thus, we neglect these
additional couplings in this letter.

 The transformation properties of all the 
matter and Higgs superfields are summarized in Table \ref{tab:fc}.
The relevant $R$-parity\footnote{More precisely, an effective $R$-parity is implemented by means of an extra $Z_2$ matter parity.}
 conserving superpotential is given
by
\begin{eqnarray}
W & = &  Y_u \hat{u^c}\hat{Q}\hat{H}_u - Y_d \hat{d^c}\hat{Q}\hat{H}_d
     + Y_{\nu}\hat{\nu^c}\hat{L}\hat{H}_u- Y_e \hat{e^c}\hat{L}\hat{H}_d
      +\mu\hat{H}_u\hat{H}_d- \mu_{R}\hat{\bar{\chi}}_R\hat{\chi}_R
     +Y_s\hat{\nu^c}\hat{\chi}_R \hat{S}
\label{eq:superpot}
\end{eqnarray} 
where $Y_e$, $Y_d$ and $Y_u$ are the usual MSSM Yukawa couplings for
the charged leptons and the quarks. In addition there are the neutrino
Yukawa couplings $Y_\nu$ and $Y_s$; the latter mixes the $\nu^c$ fields
with the $S$ fields giving rise to an inverse seesaw mechanism
for neutrino masses. For completeness we note that for realistic
neutrino masses and mixing angles
one needs also a $\mu_S \hat{S}\hat{S}$ term with a small
parameter $\mu_S$ which, however, hardly affects the Higgs sector
and, thus, is omitted here for simplicity. Its effect for the phenomenology
will be discussed elsewhere \cite{followup}. 
Note that, besides the role it plays in neutrino physics, the $Y_s$ coupling is relevant also for the Higgs  phenomenology at the loop level as it enters the mixing of $\chi_R$ and $\bar{\chi}_R$ Higgs fields with the $SU(2)_L$  Higgs doublets.  The fields $\chi_{R}$ and $\bar{\chi}_{R}$ can be viewed as the 
(electric charge neutral) remnants of $SU(2)_R$ doublets, 
which remain light in the spectrum when the $SU(2)_R$ gauge factor 
is broken by the VEV of a $B-L$ neutral triplet down to 
the $U(1)_R$.

Following the notation and conventions of \cite{Allanach:2008qq} the soft SUSY breaking Lagrangian reads
\begin{eqnarray}
V_{soft} &  =& \sum_a M_a \tilde G_a \tilde G_a +
\sum_{ij} m^2_{ij} \phi^*_i \phi_j+
 T_u \tilde{u}_R^* \tilde{Q} H_u 
- T_d \tilde{d}^*_R \tilde{Q} H_d
 + T_{\nu} \tilde{\nu}^*_R \tilde{L} H_u \nonumber \\
&&- T_e \tilde{e}^*_R \tilde{L} H_d
      +B_\mu H_u H_d- B_{\mu_{R}} \bar{\chi}_R \chi_R
     +T_s\tilde{\nu}^*_R \chi_R \tilde{S}\,.
\label{eq:soft}
\end{eqnarray} 
The first sum runs over all gauginos  for the
different gauge groups and the second
one contains the scalar masses squared.

The $U(1)_{R}\times U(1)_{B-L}$ gauge symmetry is spontaneously
broken to the hypercharge $U(1)_{Y}$ by the VEVs $v_{\chi_{R}}$ and
$v_{\bar\chi_{R}}$ of the scalar components of the $\hat\chi_R$ and
$\hat{\bar{\chi}}_R$ superfields while the subsequent
$SU(2)_{L}\otimes U(1)_{Y}\to U(1)_{Q}$ is governed by the VEVs
$v_{d}$ and $v_{u}$ of the neutral scalar components of the $SU(2)_L$
Higgs doublets $\hat H_d$ and $\hat H_u$. Thus, one can write
\begin{eqnarray}
\chi_R &=&
 \frac{1}{\sqrt{2}} \left( \sigma_R+ i \varphi_R+ v_{\chi_R}\right)
\,\,,\,\,
\bar{\chi}_R = \frac{1}{\sqrt{2}} \left( \bar{\sigma}_R 
+ i \bar{\varphi}_R + v_{\bar\chi_R}\right)\,,\\
H^0_d &=& \frac{1}{\sqrt{2}} \left( \sigma_d + i \varphi_d + v_d \right)
\,\,,\,\,\,\,\,\,\,
H^0_u = \frac{1}{\sqrt{2}} \left( \sigma_u + i \varphi_u + v_u \right)\,.
\end{eqnarray}
where the generic symbols $\sigma$ and $\varphi$ denote the CP-even and CP-odd components of the relevant fields, respectively.
 
Let us mention at this point that in order to avoid a decoupling of the 
beyond-MSSM gauge and Higgs sectors, one has to assume that the $U(1)_{R}\times 
U(1)_{B-L}$ breaking VEVs $v_{\chi_{R}}$ and  $v_{\bar\chi_{R}}$ are not very far 
from the electroweak scale. This, however, facilitates the simplified approach to 
the Higgs sector analysis where the desired 
$SU(2)_L \times U(1)_R \times U(1)_{B-L} \to U(1)_{QED}$ transition is 
treated as a one-step breaking.

At the tree level we find that in the $(\varphi_d,\varphi_u,\bar{\varphi}_R,\varphi_R)$ basis the pseudoscalar sector has  
 a block-diagonal form\footnote{Note that this remains to be the case even if 
the kinetic mixing effects are turned on.}
\begin{eqnarray}
M^2_{AA} = 
\left( \begin{array}{cc}
M^2_{AA,L} & 0 \\ 0 & M^2_{AA,R}
\end{array} \right)
\end{eqnarray}
with
\begin{eqnarray}
M^2_{AA,L} = B_\mu \left( \begin{array}{cc}
\tan \beta & 1 \\ 1 & \cot \beta
\end{array} \right) \,\,, \,\,
M^2_{AA,R} = B_{\mu_R} \left( \begin{array}{cc}
\tan \beta_R & 1 \\ 1 & \cot \beta_R
\end{array} \right) \,, 
\end{eqnarray}
 $\tan\beta= v_u/v_d$ and $\tan\beta_R= v_{\chi_R}/v_{\bar\chi_R}$. 
From these four states two are Goldstone bosons which become
the longitudinal parts of the massive neutral vector bosons $Z$ and a
$Z'$. In the physical
spectrum there are two pseudoscalars $A^0$ and $A^0_R$
with masses
\begin{equation}
 m_A^2 = B_{\mu} (\tan\beta + 1/\tan\beta) \,,\qquad
 m_{A_R}^2 = B_{\mu_R} (\tan\beta_R + 1/\tan\beta_R)
\end{equation} 
where the first formula is identical to the MSSM case.
For later convenience we define 
\begin{equation}
v^2_R = v^2_{\chi_R} + v^2_{\bar\chi_R} \,\,,\,\,
v^2 = v^2_d + v^2_u\,.
\end{equation}
The tree-level CP-even Higgs mass matrix  in the $(\sigma_d,\sigma_u,\bar{\sigma}_R,\sigma_R)$ basis 
reads 
\begin{eqnarray}\label{HHmat}
M_{hh}^2 &=&
\left(
\begin{array}{cc}
m_{LL}^2 & m_{LR}^2 \\
m_{LR}^{2,T} & m_{RR}^2
\end{array}
\right)\;,
\label{eq:Hmass}
\end{eqnarray}
where
\begin{eqnarray}\label{HHmat2}
m_{LL}^2 &=&
\left(
\begin{array}{cc}
g_Z^2 v^2 c^2_{\beta} +m_A^2 s^2_{\beta } 
& -\frac{1}{2} \left( m_A^2+g_Z^2 v^2\right) s_{2 \beta} \\ 
 -\frac{1}{2} \left( m_A^2+g_Z^2 v^2\right) s_{2 \beta} 
& g_Z^2 v^2 s^2_{\beta} +m_A^2 c^2_{\beta }
\end{array}
\right) 
\;,\\ 
m_{LR}^2 &=&
\left(
\begin{array}{cc}
 g_R^2 v v_R c_{\beta} c_{\beta_R} 
&-  g_R^2 v v_R c_{\beta} s_{\beta_R} \\ 
 - g_R^2 v v_R s_{\beta} c_{\beta_R} 
&  g_R^2 v v_R s_{\beta} s_{\beta_R} 
\end{array}
\right)\;, \\ 
m_{RR}^2 &=&
\left(
\begin{array}{cc}
 g_{Z_R}^2 v_R^2 c^2_{\beta_R} +m_{A_R}^2 s^2_{\beta_R} 
& -\frac{1}{2} \left( m_{A_R}^2+g_{Z_R}^2 v_R^2\right) s_{2 \beta_R} \\
-\frac{1}{2} \left( m_{A_R}^2+g_{Z_R}^2 v_R^2\right) s_{2 \beta_R}
&  g_{Z_R}^2 v_R^2 s^2_{\beta_R} +m_{A_R}^2 c^2_{\beta_R} 
\end{array}
\right)\;,  
\end{eqnarray}
$s_x = \sin(x)$, $c_x=\cos(x)$ ($x=\beta, \beta_R, 2 \beta, 2
\beta_R$), $g_Z^2 = (g_L^2 +g_R^2)/4$, $g_{Z_R}^2 = (g_{BL}^2
+g_R^2)/4$ and $g_{L}$, $g_{R}$ and $g_{BL}$ are gauge couplings
associated to the $SU(2)_L$, $U(1)_R$ and $U(1)_{B-L}$ gauge factors,
respectively.  The matrix $m_{LL}^2$ contains the standard MSSM
doublet mass matrix\footnote{To see this explicitly one has to
integrate out the additional Higgs fields in the $v_R \to \infty$
limit which yields a shift in the gauge couplings such the the MSSM
limit is achieved.}, $m_{RR}^2$ corresponds to the $U(1)_{R}\times
U(1)_{B-L}$ Higgs bosons and $m_{LR}^2$ provides the essential mixing
among the two sectors. It is in particular this sector which gives
rise to the increase of the mass of the MSSM-like lighter Higgs boson
already at tree-level overcoming the stringent MSSM bound.  In what
follows we shall denote the orthogonal matrix diagonalizing the mass
matrix in \eq{Hmass} by $R$ and the eigenvalues/eigenstates will be
ordered in such a way that $m_i \le m_j$ for $i<j$.  In a full analogy
to the MSSM, the entire Higgs spectrum can be parametrized in terms of
the pseudoscalar masses $m_{A}$ and $m_{A_{R}}$ and the relevant
mixings encoded by $\tan\beta$ and $\tan\beta_R$.

The similarity to the MSSM makes it also clear that the loop
corrections can be potentially large and, thus, very important.  In
particular, top and stop loops affect the $SU(2)_L$-doublet part of
the mass matrix in the usual manner. Furthermore, in certain parts of
the parameter space also the neutrino/sneutrino loops can be large.
Technically, we have been using the {\tt SARAH} package
\cite{Staub:2008uz,Staub:2010jh} to obtain the relevant
$SU(2)_{L}\otimes U(1)_{R}\otimes U(1)_{B-L}$ generalizations of the
basic MSSM formulae given in \cite{Pierce:1996zz}; in this respect, let us stress that this accounts for the full one-loop structure of the Higgs sector. For further details
an interested reader should refer to a dedicated work
\cite{followup}.

Finally, besides direct bounds from various Higgs boson searches,
there is an important constraint on the parametric space of the model
associated to the heavy $Z'$.  In the $(W^0_3, B, B')$ basis
(corresponding to the electrically neutral generators of $SU(2)_L$,
$U(1)_R$ and $U(1)_{B-L}$, respectively) the relevant vector boson
mass matrix reads
\begin{eqnarray}
M^2_{VV} &=& 
\frac{1}{4} \left( \begin{array}{ccc}
  g_L^2 v^2 &
 0                 & 
 -  g_L g_R v^2                      \\
0                 &
    g_{BL}^2 v_R^2                   &
    - g_{BL} g_R v_R^2                                           \\
 -  g_L g_R v^2                      &
    - g_{BL} g_R v_R^2                                           &
     g_R^2 \left( v^2 + v_R^2 \right)
\end{array} \right)\,,
\end{eqnarray}
from where the masses of the photon, $Z$ and $Z'$ are readily identified
\begin{equation}
m_\gamma = 0\,,\quad
m_Z^2 = 
  \tfrac{1}{8} \left( A - \sqrt{A^2- 4 B} \right) \,,\quad
m_{Z'}^2 =
  \tfrac{1}{8} \left( A + \sqrt{A^2- 4 B } \right) \,,
\end{equation}
where
\begin{equation}
A = 
   \left(g_L^2+g_R^2\right) v^2 + \left(g_{BL}^2 + g_R^2\right)  v_R^2 \,,\quad
B =
  \left[ g_L^2 \left(g_R^2 + g_{BL}^2 \right) + g_{BL}^2 g_R^2 \right] 
   v^2 v_R^2 \,.
\end{equation}
In particular, the product $g^2_{Z_R} v^2_R \,[= m^2_{Z'} +
 O(v^2/v^2_R)]$ is constrained by the $Z'$ searches at LEP and at
 Tevatron as well as from the precision measurements
 \cite{Nakamura:2010zzi,Erler:2011ud}.

\section{Numerical results}
\label{sec:numerics}

The numerical results given below have been calculated in {\tt
SPheno} \cite{Porod:2003um,Porod:2011nf} for which the necessary 
subroutines and input
files were generated by the relevant extension of {\tt SARAH}
\cite{Staub:2011dp}.  Hence, the complete one-loop corrections in the
extended Higgs sector have been included \cite{followup}. 
 We will concentrate the discussion on the lightest two 
mass eigenstates, since here the changes with respect to the 
MSSM are expected to be most important for the choice of 
$m_A$ and $m_{A_R}$ used below. We always check that we are at the minimum of the potential, 
by solving the (1-loop improved) tadpole equations 
for the soft Higgs masses.

Throughout the numerical analysis we have adopted a CMSSM-like configuration specified
by $M_{1/2}= 600$~GeV, $m_{0}= 120$~GeV, $A_{0}= 0$ and 
$\tan\beta=10$. The stop-sector soft masses in (\ref{eq:soft}) were chosen 
as  
$m_{\tilde Q_3} = m_{\tilde U_3} = 2$~TeV, $T_{u33}=3$~TeV 
and the top quark mass has been fixed to $m_t= 172.9$~GeV.  
In addition we have assumed $v_{R}=5$ TeV, $\mu=800$~GeV,
$m_A=800$~GeV, $\mu_\chi=-500$~GeV, $m_{A_R}=2$~TeV and
$\tan\beta_R=1.1$ unless specified otherwise\footnote{
For this choice of parameters $m_{h^0}$ in the MSSM limit 
is about 125 GeV (1-loop), while for $m_{\tilde{Q}_{3}}=1.1$~TeV, $m_{\tilde{U}_{3}}=0.96$~TeV and $T_{u33}=1.1$~TeV corresponding to the RGE 
solutions for these CMSSM one finds $m_{h^0}=111$ GeV at 
1-loop.}. For the sake of completeness\footnote{It is perhaps worth mentioning that from the effective theory point of view the specific values of $g_{BL}$ and $g_{R}$ do not matter as long as they yield the correct MSSM hypercharge coupling. Indeed, we have verified that different choices lead to results very similar to those quoted in the text.}
we have taken $g_{BL}=0.46$ and $g_R=0.48$.

Some further remarks concerning the parameters of the extended Higgs sector are
in order here. The experimental constraints on the $Z'$ mass yield
a lower bound on $v_R$ of about 2.5~TeV 
\cite{Erler:2011ud}\footnote{Our $Z'$ corresponds to the $Z_\chi$ in
the notation of \cite{Erler:2011ud}.} for the assumed gauge couplings. 
This VEV, however, also enters
the sfermion mass matrices via the $D$-term contributions. Focusing, e.g.,
at the charged sleptons the relevant mass matrix reads
\begin{eqnarray}
M^2_{\tilde l} = \left( \begin{array}{cc}
 M^2_{\tilde L} + \frac{1}{8} M_{DL}^2 + m_f^2 & 
 \frac{1}{\sqrt{2}} \left( v_d T_l - \mu Y_l v_u \right) \\
 \frac{1}{\sqrt{2}} \left( v_d T_l - \mu Y_l v_u \right) &
 M^2_{\tilde E} + \frac{1}{8} M_{DR}^2 + m_f^2
\end{array} \right)\,,
\end{eqnarray}
where
\begin{equation}
M^2_{DL} = 
 g_{BL}^2 ( v^2_{\chi_R} - v^2_{\bar\chi_R} ) + 
 g_L^2 ( v_u^2 - v_d^2 )\quad\text{and}\quad
M^2_{DR} =
(g_{R}^{2} - g_{BL}^2) ( v^2_{\chi_R} - v^2_{\bar\chi_R} ) + 
   g_R^2 ( v_u^2 - v_d^2 )
\end{equation}
are just the $D$-terms, 
$M^2_{\tilde L}$ and $M^2_{\tilde E}$ are the soft SUSY masses
for the L-type and R-type sleptons and all flavour indices have been suppressed in the above formula.

Since the (dominant) $v_R^2$-parts of the two $D$-terms have opposite
signs, the breaking of
the extra gauge group must be nearly ``$D$-flat'', i.e., $\tan\beta_R\simeq 1$
as otherwise one of the sleptons would become tachyonic\footnote{Let us note that this is indeed the case in all supersymmetric models featuring a spontaneously broken extended gauge symmetry well above the TeV scale (like, e.g., SUSY GUTs) and as such this requirement should be viewed as a phenomenological constraint rather than a fine-tuning.}. For
completeness we also note that the cases of $\tan\beta=1$ and $\tan\beta_R =1$
lead to saddle points of the potential but not to minima which is a well known
fact within the MSSM. In a complete analogy with
the MSSM  one can also show that for  $\tan\beta_R  \to 1$ one of the Higgs states gets
massless at the tree level. 
Thus, since $\tan\beta_R$ has to be close to one, we generally expect 
 two light Higgs bosons in the spectrum, which holds even at the one-loop level.

\begin{figure}
 \begin{center}
  \hspace{-20mm}
  \begin{tabular}{cc}
   \includegraphics[height=5cm]{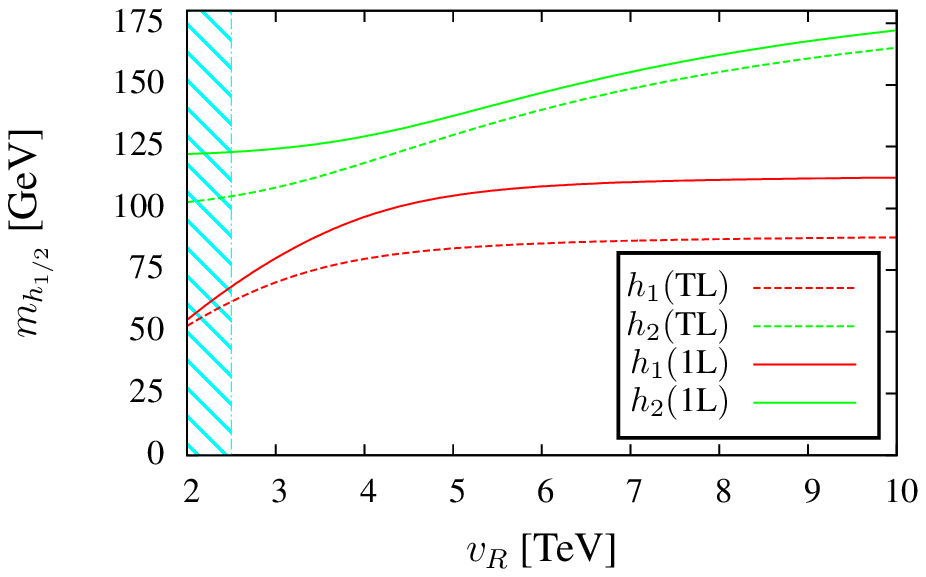}
   \includegraphics[height=5cm]{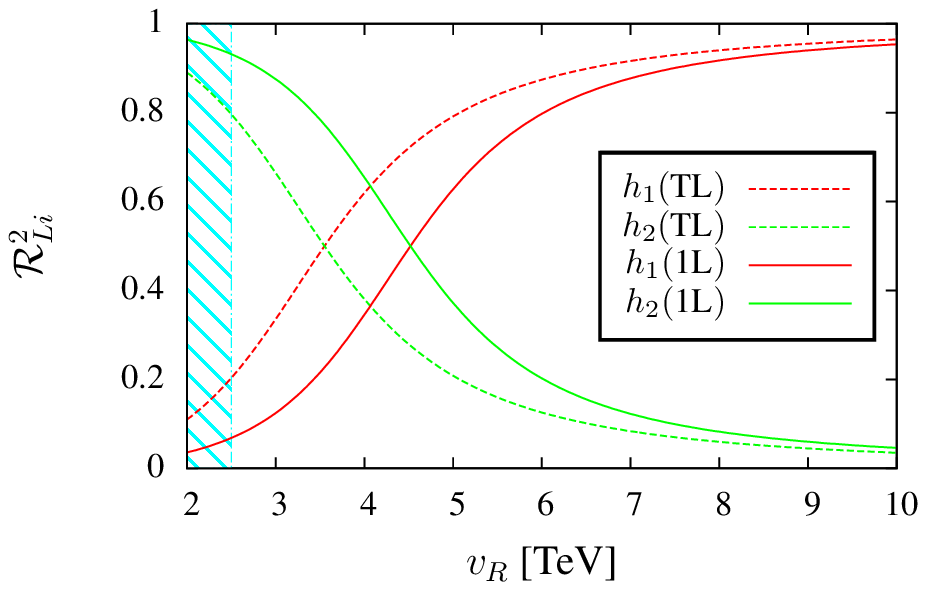}    
   \end{tabular}
   \caption{\label{fig:higgsvsvR_TLloop} The tree level and one-loop 
    masses of the two lightest Higgs bosons $h_{1,2}$ (left) and ${\cal R}^2_{Li}$
     (right) as a function of $v_R$; at tree level (TL) in dashed and at one loop (1L) in solid lines.
    The values of all the other parameters are given in the text. The shaded area
    is excluded by the $Z'$ searches.}
 \end{center}
\end{figure}

In Figure~\ref{fig:higgsvsvR_TLloop}
we show the masses of the two lightest Higgs bosons together with 
\begin{equation}{\cal R}^2_{Li} \equiv R^2_{i1} + R^2_{i2}\end{equation} as a function of $v_R$ where $i=1,2$ labels the light Higgs scalars in the model. 
Note that the quantity
${\cal R}^2_{Li}$, which reaches one in the MSSM limit, is a rough measure of 
how much the corresponding Higgs with index
$i$ resembles an MSSM Higgs boson.
 Roughly speaking, the smaller this
quantities is, the smaller is the $i$-th Higgs coupling to the $Z$- and $W$-bosons, implying a
reduced production cross sections at LEP, Tevatron and the LHC.

As claimed above there are two light CP-even states $h_{1,2}$ which essentially correspond to 
an admixture of the ``standard'' MSSM-like doublet component $h^0$ and its counterpart $h^0_R$ spanning over the 
$\chi_R-\bar{\chi}_R$ sector; this can also be seen
by noticing that ${\cal R}^2_{L1} + {\cal R}^2_{L2} \simeq 1$ 
as displayed on the right hand
side of Figure~\ref{fig:higgsvsvR_TLloop}. We stress that  the state which mainly
resembles the MSSM $h^0$ (i.e., the one with a large ${\cal R}^2_{Li}$) has already a tree-level mass of around 110 GeV or larger and reaches  up to 140 GeV once loop corrections are 
included\footnote{Actually, even larger values can be obtained when varying the 
parameters, e.g.~$m_{A_R}$.}.
The lighter state with a mass below 100 GeV hardly couples to the
$Z$-boson and, thus, the LEP constraints from the Higgs searches do not
apply for it. To this end, we have used the {\tt HiggsBounds} package \cite{Bechtle:2008jh,Bechtle:2011sb} to check explicitly that all the configurations of our concern here are experimentally allowed. Note also that the
large variation in ${\cal R}^2_{Li}$  as seen on the right panel of the
Figure \ref{fig:higgsvsvR_TLloop} and, in particular, its high sensitivity to radiative corrections is expected because the parameters have been deliberately 
chosen close to a level-crossing region.

\begin{figure}
 \begin{center}
  \hspace{-12mm}
  \begin{tabular}{cc}
   \includegraphics[height=5cm]{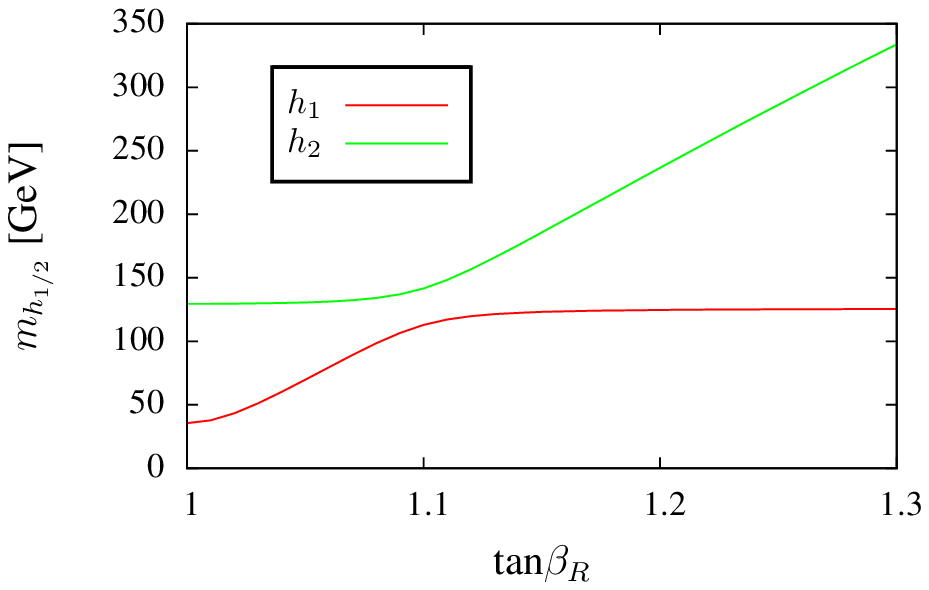} &
   \includegraphics[height=5cm]{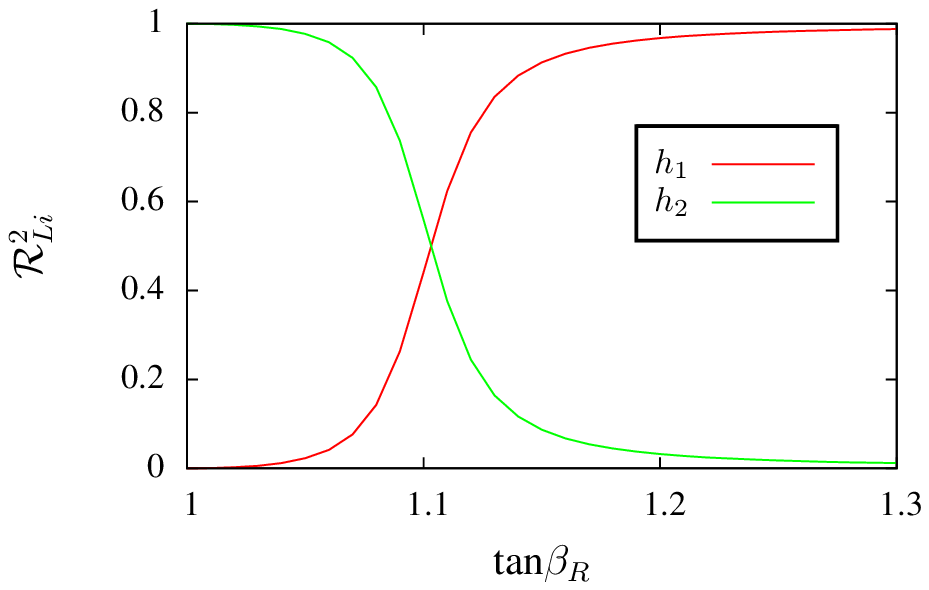} \\
   \includegraphics[height=5cm]{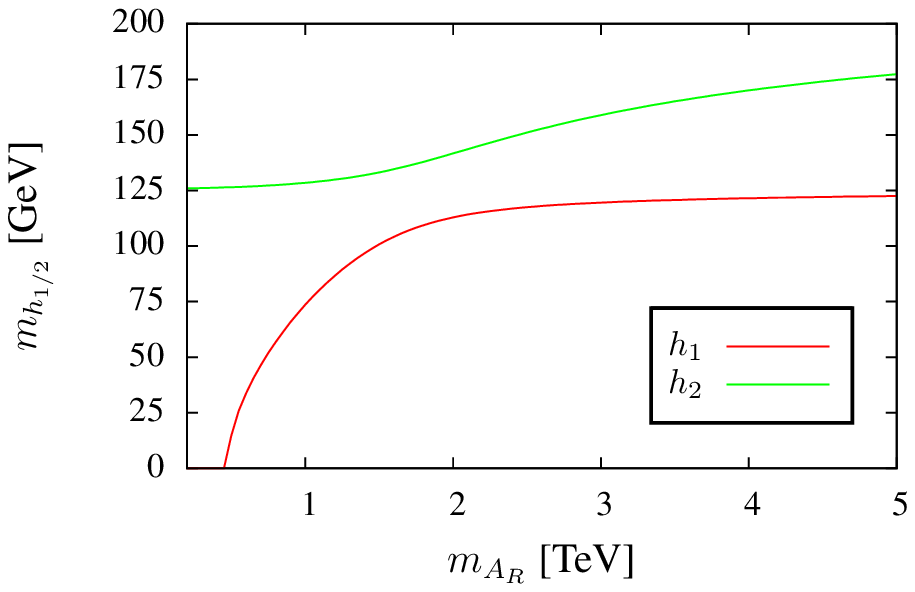}  & 
   \includegraphics[height=5cm]{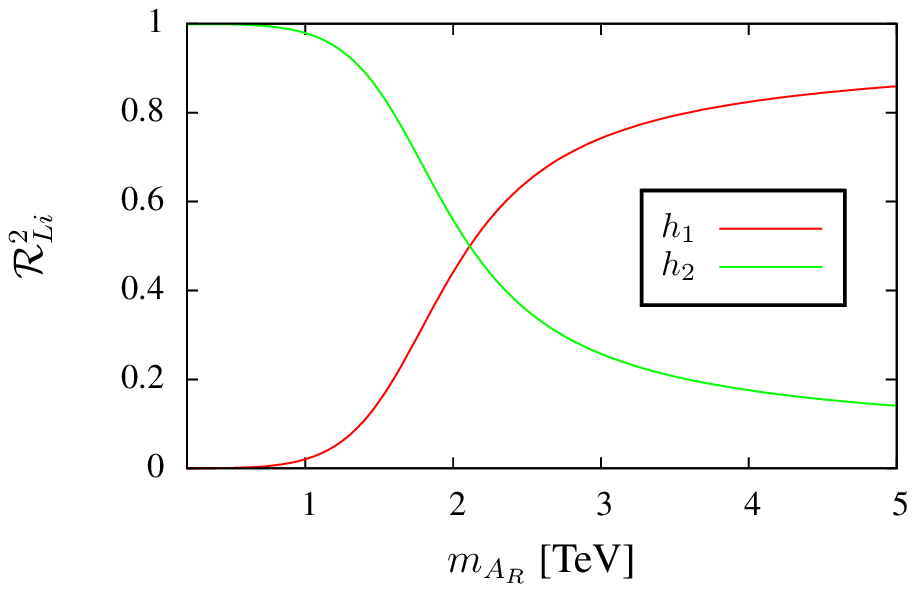} 
  \end{tabular}
   \caption{\label{fig:vRtanbRmARvshhANDlh_1}
    One-loop masses of the two lightest Higgs bosons (left column) and
    ${\cal R}^2_{Li}$ (right column) as a function of
    $\text{tan}\beta_R$ (upper row) and $m_{A_R}$ (lower row). The values of all the other parameters are given in the text.}
 \end{center}
\end{figure}

This can be also seen in Figure \ref{fig:vRtanbRmARvshhANDlh_1} where
we display the $m_{h_{1,2}}$ dependence on $\text{tan}\beta_R$ and
$m_{A_R}$. All results shown in this figure are at the one-loop level.
The upper bound on $\text{tan}\beta_R$ is given by the requirement
that for a given value of $v_R$ all sfermions masses are consistent
with existing data (which, however, depends also on the sfermion mass
parameters).  The observed dependence on $\text{tan}\beta_R$ is,
indeed, rather strong. Note also that very light $h_{1}$ can be
obtained for\footnote{The exact value as well as the others given
below depend on the other parameters.} $\text{tan}\beta_R \lsim 1.05$.
As in this regime it is mainly a combination of $\bar{\chi}_R$ and
$\chi_R$ (see the right panel) the usual bounds do not apply. However,
the second lightest Higgs boson (similar to the MSSM $h^0$) can decay
into a pair of these states with sizable branching ratio which in turn 
can change the 
Higgs phenomenology drastically~\cite{followup}.
For $1.2 \lsim\text{tan}\beta_R \lsim 1.3$ the lightest
state becomes mainly  the MSSM $h^0$ with a mass close to 130 GeV
which is a consequence of the stop-sector parameter choice. 
In this figure one also sees that there is still quite some mixing
between the two lightest states even for $m_{A_R} = 5$~TeV; this implies a 
change in the phenomenology with respect to that of the MSSM (for a given set 
of the MSSM parameters).

\begin{figure}
 \begin{center}
  \hspace{-20mm}
  \begin{tabular}{c}
   \includegraphics[height=5.5cm]{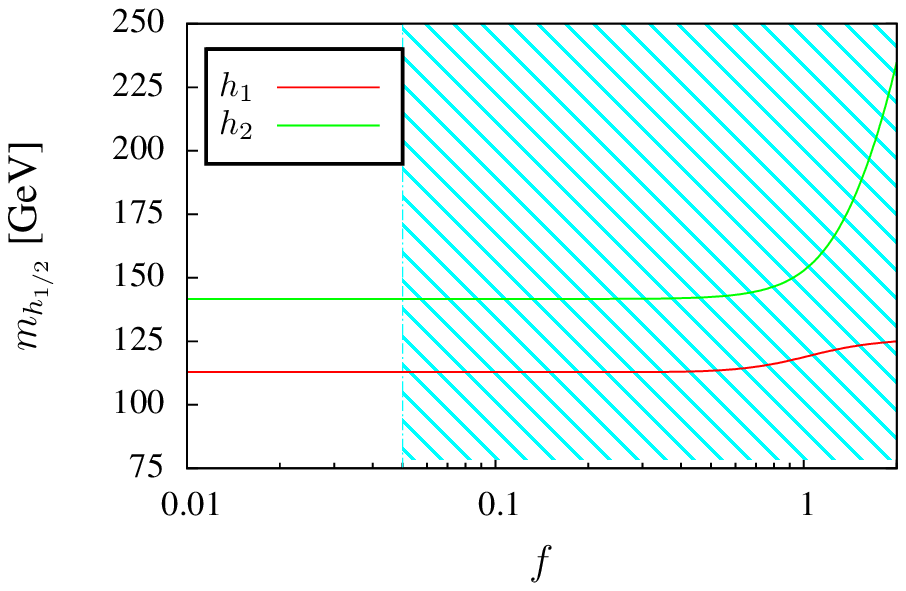}
  \end{tabular}
   \caption{\label{fig:fyuk_1} One-loop masses of the two lightest Higgs bosons 
    as functions of $f$. The values of all the other parameters are given in 
    the text. 
    The shaded area is excluded by $\mu\to e\gamma$.}
 \end{center}
\end{figure}
We checked that in this model, in general, the loops due to third generation
sfermions (in particular the stops) give the largest contribution. 
In reference \cite{Elsayed:2011de} it has been shown that in inverse
seesaw models also the sneutrino loops can give large contributions.
Indeed, we find
that there can be huge contributions if the neutrino Yukawa couplings
are $O(1)$ or larger as can be seen in Figure \ref{fig:fyuk_1}.
 The neutrino Yukawa couplings
are parametrized as
\begin{equation}
Y_\nu = f \left( \begin{array}{ccc}
 0 &   0 &  0 \\
 a &   a &  -a \\
 0 &  1 &  1 \\
\end{array}  \right) \,,
\end{equation}
with
\begin{equation}
a = \left( {\Ds}/{\Da} \right)^{\frac{1}{4}} \sim 0.4\;,
\end{equation}
and the structure has been chosen such that one correctly 
accommodates the neutrino data.
 However, we find that the bound 
BR$(\mu \to e \gamma) \lsim 2.4 \cdot 10^{-12}$ \cite{Adam:2011ch}
severely constrains this option as, for large $f$, one gets a large contribution to $\mu\to e\gamma$
due to the chargino-sneutrino and $W$-neutrino loops. 

There are,
of course, several ways to tune the parameters such that this
bound is avoided. For example, one can add a non-minimal flavour
structure into the slepton sector \cite{Bartl:2003ju} or tune
the structure of the neutrino Yukawa couplings so that very
specific values for $\theta_{13}$, the reactor mixing angle, are obtained
\cite{Antusch:2006vw,Esteves:2010ff}. This implies that, in principle,
larger values for the neutrino Yukawa couplings are possible, hence rendering
the corresponding loops more important. On the other hand,  making
them competitive even to the stop loops already requires quite some 
tuning \cite{followup}.

\section{Conclusions}
\label{sec:conc}

In this letter we have discussed the Higgs sector of a 
supersymmetric model
where the SM gauge group has been extended to 
$SU(3)_{c}\times SU(2)_L\times U(1)_R \times U(1)_{B-L}$. In particular, we have shown that, already at
the tree-level, the CP-even Higgs boson resembling the lightest neutral Higgs $h^0$ of the MSSM,
can have a mass well above $m_Z$. At the one loop level, masses of
 140 GeV and even above can easily be reached. In addition to such an $h^0$-like Higgs, one can also have a second light state which, however,
hardly couples to the SM vector bosons as it predominantly spans over the  
SM-neutral components. We have found regions where the $h^0$-like Higgs
can decay into two such states which, however, alters the standard
search techniques at the LHC.
Finally, we would like to stress that the general features discussed
here also apply to other extensions of the SM gauge group, e.g., to full-featured
left-right symmetric models, provided the MSSM Higgs doublets are
charged with respect to the extended gauge symmetry.

\section*{Acknowledgements}

W.P.~thanks the IFIC for hospitality during an extended stay
and the Alexander von Humboldt foundation for financial support.
F.S.\ and W.P.\  have been supported by the DFG, project
number PO-1337/1-1.~M. M. is supported by the Marie Curie Intra European Fellowship within the 7th European Community Framework Programme FP7-PEOPLE-2009-IEF, contract number PIEF-GA-2009-253119.
We acknowledge support from the Spanish MICINN grants FPA2008-00319/FPA,
FPA2011-22975 and MULTIDARK CSD2009-00064 and
by the Generalitat Valenciana grant Prometeo/2009/091 and the
EU~Network grant UNILHC PITN-GA-2009-237920.

\end{document}